\documentclass[conference]{IEEEtran}

\AtBeginDocument{%
  \providecommand\BibTeX{{%
    \normalfont B\kern-0.5em{\scshape i\kern-0.25em b}\kern-0.8em\TeX}}}

\newtheorem{theorem}{Theorem}
\newtheorem{definition}{Definition}

\newcommand{\JS}[1]{\textcolor{magenta}{#1}}

\newcommand{\Lg}{\mathcal{L}}
\newcommand{\T}{\mathcal{T}}
\newcommand{\A}{\mathcal{A}}
\newcommand{\M}{\mathcal{M}}
\newcommand{\Pl}{\mathcal{P}}
\newcommand{\St}{\mathcal{S}}
\newcommand{\F}{\mathcal{F}}
\newcommand{\D}{\mathcal{D}}
\newcommand{\V}{\mathcal{V}}

\newcommand{\Sg}{\text{StateGen}}
\newcommand{\Sv}{\text{StateVer}}
\newcommand{\Pg}{\text{PGen}}
\newcommand{\Pv}{\text{PVer}}
\newcommand{\Vg}{\text{VGen}}
\newcommand{\Vv}{\text{VVer}}
\newcommand{\WPg}{\text{WPGen}}
\newcommand{\WPv}{\text{WPVer}}
\newcommand{\W}{\mathcal{W}}
\newcommand{\Conf}{\text{Conf}}

\newcommand{\myparagraph}[1]{\par\smallskip \noindent{\bf #1} }

\usepackage{enumerate}
\usepackage[colorinlistoftodos]{todonotes}
\usepackage[normalem]{ulem}
\usepackage{graphicx}
\usepackage{subfig}
\usepackage{algorithm}
\usepackage{amsmath}
\usepackage{algpseudocode}
\usepackage[hyphens,spaces,obeyspaces]{url}
\usepackage[section]{placeins}

\makeatletter
\def\ALG@special@indent{%
    \ifdim\ALG@thistlm=0pt\relax
        \hskip-\leftmargin
    \else
        \hskip\ALG@thistlm
    \fi
}
\newcommand{\Input}[1]{\item[]\noindent\ALG@special@indent \textbf{Input:}\ #1}
\newcommand{\Output}[1]{\item[]\noindent\ALG@special@indent \textbf{Output:}\ #1}

\newcommand{\Verify}[1]{\textbf{verify:}\textit{\ #1}}

\makeatletter
\def\blfootnote{\gdef\@thefnmark{}\@footnotetext}
\makeatother

\begin{document}              

\title{Verifiable Observation of Permissioned Ledgers}

\author{
  \IEEEauthorblockN{
    Ermyas Abebe\IEEEauthorrefmark{1},
    Yining Hu\IEEEauthorrefmark{1},
    Allison Irvin\IEEEauthorrefmark{1},
    Dileban Karunamoorthy\IEEEauthorrefmark{1},
    Vinayaka Pandit\IEEEauthorrefmark{1},\\
    Venkatraman Ramakrishna\IEEEauthorrefmark{1},
    Jiangshan Yu\IEEEauthorrefmark{2}$^\P$
  }
  \IEEEauthorblockA{
    \IEEEauthorrefmark{1}IBM Research\\
    \IEEEauthorrefmark{2}Monash University}\\
}
\maketitle

\blfootnote{$^{\P}$ Corresponding author. Email: jiangshan.yu@monash.edu}
\begin{abstract}
Permissioned ledger technologies have gained significant
traction over the last few years. 
For practical reasons, their applications have focused
on transforming narrowly scoped use-cases in isolation. This has led
to a proliferation of niche, isolated networks that are quickly
becoming data and value silos. To increase value across the broader
ecosystem, these networks must seamlessly integrate with existing systems
and interoperate with one another. A fundamental requirement for
enabling cross-chain communication is the ability to prove the
validity of the internal state of a ledger to an external
party.  However, due to the closed nature of permissioned ledgers, their
internal state is opaque to an external observer. This makes consuming and verifying states from these networks a non-trivial problem.

This paper addresses this fundamental requirement for state sharing
across permissioned ledgers. In particular, we address two key
problems for external clients: (i) assurances on the validity of state
in a permissioned ledger and (ii) the ability to reason about the
currency of state. We assume an adversarial model where the members of
the committee managing the permissioned ledger can be malicious in the
absence of detectability and accountability. We present a
formalization of the problem for state sharing and examine its
security properties under different adversarial conditions. We propose
the design of a protocol that uses a secure public ledger for
providing guarantees on safety and the ability to reason about time,
with at least one honest member in the committee. We then provide a
formal security analysis of our design and a proof of concept
implementation based on Hyperledger Fabric demonstrating the effectiveness of the proposed protocol.








\end{abstract}

\graphicspath{{./figures/}}


\section{Introduction}\label{SEC:introduction}

Permissioned networks are designed to address the needs of enterprises
collaborating across organizational boundaries using a shared
ledger~\cite{Fabric,Corda,Besu,Quorum,MultiChain}. They restrict
membership to a known group of identities to keep data private and
confidential. Unlike permissionless networks, the security in these
systems relies on the ability to detect and hold parties accountable
for their actions. The use of a shared ledger between enterprises can
reduce counter-party risk and mitigate the need for costly and
time-consuming dispute resolution processes, which often involves
legal and judicial systems.

For practical reasons, the adoption of permissioned ledgers has thus
far been driven through use-cases. Enterprises have been coalescing
into consortia to create networks that address narrow use-cases in
isolation. These use-cases often represent one part of a complex
business process. For example, various networks have emerged that
focus on provenance~\cite{IFT,Everledger,Ripe}, trade
logistics~\cite{TradeLens} and
trade-finance~\cite{WeTrade,Komgo,MarcoPolo}, all of which address
different aspects of the broader process of global trade. This has led
to a proliferation of niche and isolated networks that are quickly
becoming data and value silos. To increase value across the broader
ecosystem, permissioned networks must be able to securely communicate
with other networks while integrating with existing legacy
applications.

However, the restrictive nature of a permissioned network introduces
two key challenges to an external observer: (i) the inability to
verify that state shared externally is the same as the state agreed on
internally and (ii) the inability to reason about the currency of the
state as a result of having no visibility on the chain's progress. An
ostensible solution to these problems is for the external client to
query one or more parties in the network and obtain a set of notarized
responses~\cite{Abebe2019,Cactus2020}. However, in the absence of a
system for detectability and accountability outside of the network, a
malicious committee can choose to misrepresent facts by either
presenting false or stale state.

This paper addresses the problem of state sharing across permissioned
networks. We assume an adversarial model where all members of a
committee can be malicious except for a single honest party. We rely
on the fact that a single honest member can hold parties accountable
if it is able to detect any misrepresentation of facts with external
clients. In our approach, the permissioned network publishes
commitments representing snapshots of its internal state at fixed
intervals to a trusted third-party such as a secure public ledger.
These 
commitments enable both honest members to verify state snapshots while
raising disputes when conflicts are detected, and external clients to
verify proofs against valid commitments according to a common notion
of time. The process of resolving disputes---which in practice can take
various forms such as fines, termination of membership in the network,
or legal proceedings---is out of scope of this paper.




\myparagraph{Related Work.} The problem of a client needing to verify
the state of a blockchain without directly observing it is not
new. Several light client
implementations~\cite{PoPoW,kiayias2017non,nakamoto2019bitcoin}
provide efficient ways to verify the inclusion of transactions in a
ledger assuming partial access to the ledger data, such as block
headers. More recently, several non-custodial Layer-2
protocols~\cite{gudgeon2020sok} that rely on the parent chain for
security have become popular. Examples of these systems include
Plasma~\cite{ethhubplasma}, NOCUST~\cite{khalil2018nocust}, Optimistic
Rollups schemes such Arbitrum~\cite{kalodner2018arbitrum} and
ZK-Rollups schemes such as zkSync~\cite{matterLabs,zksync}. Although
these systems allow transactions to take place off-chain, their state
is periodically check-pointed on the parent chain for
settlement. These approaches rely on external users and watch services
having access to the off-chain state in order to keep the operators
honest, using a challenge-response game. Cross-chain communication
protocols such as Cosmos~\cite{Cosmos}, Polkadot~\cite{Polkadot},
Cardano~\cite{cardano}, Ren~\cite{ren} and BTC Relay~\cite{btcrelay}
rely on a network of intermediate nodes having partial access to a
chain's data in order to ensure validity of the state communicated. In
contrast to these systems, our work assumes that an external client
has no visibility of the internals of a permissioned ledger, including
data such as block headers. A simplified
approach~\cite{Abebe2019,Cactus2020} to this problem requires an
external client placing significant trust on the committee in order to
rely on their claims. Alternatively, schemes such as
ZK-SNARKS~\cite{ben2013snarks}, ZK-STARKS~\cite{ben2018scalable} and
Bullet-Proofs~\cite{bunz2018bulletproofs} are too complex and
expensive approaches for addressing a range of general purpose uses
cases that permissioned ledgers are designed for. Furthermore, it
would be impractical to migrate today's real-world deployments to
these schemes. See Appendix~\ref{sec:rw} for further discussion.

\myparagraph{Contributions.} To summarize, this paper makes the
following contributions: (i) A formalization of the problem of state
sharing across permissioned ledgers, first introducing a formal model
of permissioned ledgers and then presenting a formal model for state
sharing, along with their security properties in different adversarial
conditions. To the best of our knowledge, this is the first attempt at
providing a formal treatment of this problem.  (ii) The design of a
protocol for state sharing based on the formal model. The protocol
uses a secure public ledger that acts as a bulletin board for state
commitments and a global clock for synchrony. We examine the
protocol's safety and liveness properties in the presence of at least
one honest member in the network. (iii) A security analysis of the
protocol proving the stated properties. (iv) A proof-of-concept
implementation and evaluation of the protocol using Hyperledger
Fabric~\cite{Fabric}.

The remainder of the paper is organized as
follows. Section~\ref{SEC:network-formalisms} formalizes state sharing
in permissioned networks and examines their security
properties. Section~\ref{SEC:protocol} proposes and discusses the
design of a protocol for state sharing that is secure against a
malicious but cautious committee. Section~\ref{SEC:proof-of-concept}
discusses a proof of concept implementation based on Hyperledger
Fabric with security analysis. Section~\ref{SEC:perf-eval} provides an
evaluation of the protocol and, finally, Section~\ref{SEC:conclusion}
offers concluding remarks.


\section{Formalizing State Sharing in Permissioned Networks}\label{SEC:network-formalisms}
%

In this section we present a formal treatment of state sharing from a
permissioned ledger to external parties. We first identify the parties
involved and the different adversary models to consider. We then
formalize essential properties of permissioned ledgers, and use these
to describe the problem of state sharing. We do this by distinguishing
between the internal state of a ledger, which is visible to parties
within a network, and external views of a ledger, which are the
observations that external parties can make. We introduce algorithms
for generating and verifying proofs about the validity of internal
state and show how these can be adapted to prove validity of state
against one or more external views, thus enabling external clients to
verify state independently. Using these constructs, we present
formally the security guarantees that state sharing mechanisms must
provide under different adversary models.

\subsection{System model}

\myparagraph{Environment.} In the communication between a
permissioned ledger and external entities, we consider three main types
of participants, namely internal clients, members of the management
committee, and external clients. The management committee
has full knowledge of the ledger and is responsible for its maintenance.
Internal clients do not have direct visibility or direct access to the ledger
but interact with the management committee to query and update states according
to defined policies. Both the management committee and internal clients are
considered participants of the permissioned network. External clients, on the
other hand, are not participants of the network, but still have a need to obtain
and verify states from the permissioned network.

\myparagraph{Adversary model.} We consider three types of adversary
scenarios with respect to the management committee of a permissioned
blockchain, as follows.
  \begin{itemize}
  \item \textbf{Trustworthy committee.} Some members of the management
    committee can be faulty. However, the committee as a whole is
    \emph{trustworthy} as it can tolerate faults when making
    agreements.

  \item \textbf{Malicious but cautious committee.} The management
    committee may be \emph{malicious but cautious}. It will
    behave arbitrarily when its malicious behaviors cannot be detected
    and are not accountable.


  \item \textbf{Malicious committee.} The management committee is
    \emph{malicious}. It will behave arbitrarily with no other
    considerations.
  \end{itemize}
 
  Internal clients in a permissioned network generally assume a
  trustworthy committee. While they trust the management committee as
  a whole, they may not trust any individual member. External clients,
  on the other hand, have different levels of trust in the management
  committee of a permissioned ledger depending on the adversary
  model. While in some cases an external client may trust the
  consensus agreement of a permissioned chain (i.e. trustworthy
  committee), in many cases such a strong trust relationship cannot be
  assumed. This is largely because external clients do not have full
  visibility into the ledger and might not have reliable affiliations
  with internal participants of the network. Conversely, as
  permissioned ledgers are often maintained by a reputable consortium,
  it is unlikely that the entire management committee of a
  permissioned ledger is fully malicious. Thus, we focus instead on
  the more practical adversary scenario: the scenario where the
  management committee is \emph{malicious but cautious}.
  
  We assume that the permissioned ledger has sufficiently strong
  internal consensus mechanisms to thwart malicious behaviors from a
  predefined threshold of members of the management committee. We also
  assume that at least one member of the management committee is
  honest in its interactions with external clients. If no such member
  exists, the management committee is considered fully malicious and
  the ledger is deemed completely unreliable.

\subsection{Modeling permissioned ledgers}\label{SEC:formal-permissioned-networks}

We define a permissioned ledger as a state machine where the
transition of states is processed by executing a sequence of
transactions. The state machine is replicated and maintained by the
management committee. The committee processes transactions and updates
system states according to defined policies.

\begin{definition}
  A permissioned ledger with $n$ state transitions is a tuple
  $\Lg^{n}=(\T_i, \A_i, \M_i, \Pl_i)_{i=1}^n$ of sets $\A_i\in\A$ of
  application states, sequences $\T_i\in \T$ of transactions,
  management committees $\M_i\in\M$, and management policies
  $\Pl_i\in \Pl$. Here, $\A$ refers to the set of all possible
  application states, $\T$ refers to the set of all possible
  subsequences of the transactions recorded on the ledger, $\M$ refers
  to the set of all possible management committees, and $\Pl$ refers
  to the set of all possible management policies. We denote by $\St_0$
  the genesis state and by $\St_n$ the current state of $\Lg^{n}$.
\end{definition}

Let $\St_i\in\St$ be the $i$-th state of $\Lg^{n}$ for some
$i\in[1,n]$. Here, $\St$ refers to the set of all possible ledger
states. Each state $\St_i$ is a tuple $(\A_i, \M_i, \Pl_i)$ of a set
$\A_i$ of application states, a management committee $\M_i$, and a
management policy $\Pl_i$. A transaction applied to state $\St_i$
updates one or more application states, the management committee, or
the management policy from its current state $\St_i$ to the resulting
state $\St_{i+1}$. We use $A_{i,j}\in\A_i$ to denote the $j$-th
application state at state $\St_i$.  Similarly, $T_{i,j}\in\T_i$
denotes the $j$-th transaction at state $\St_{i}$. We denote by
$\Lg_i\in\Lg$ the $i$-th tuple $(\T_i, \St_i)$ of $\Lg^{n}$.

The application states take the form of either (i) a set of UTXOs or (ii) a set of
key/value pairs in the case of an account model. These states are
application dependent and can represent arbitrary data or assets.  We
assume that all transactions in a ledger are totally ordered. In
practice the sequence $\T_i$ of transactions can represent a single
transaction in some protocols \cite{Corda} or a block of transactions
in others \cite{Fabric}, \cite{Besu}. The management committee $\M_i$
represents the set of members responsible for state $\St_i$ according
to the policy $\Pl_i$.

We define a pair of algorithms $(\Sg, \Sv)$. Given a state and a
sequence of transactions to be applied on this state as input, $\Sg$
outputs a new state. The correctness of this transition can be
verified by $\Sv$.

\begin{definition}\label{def:stategenver}
  State transition and verification of ledger $\Lg^{n}$ is
  accomplished by a pair of algorithms $(\Sg, \Sv)$ such that for all
  $i\in[1,n]$, we have
  \begin{itemize}
  \item $\Sg: \T \times \St \to \St$
  \item $\Sv: \T \times  \St \times  \St \to \text{True/False}$
  \end{itemize}
  and  $$\forall i\in[1,n],\ \Sv(\T_i, \St_{i-1}, \Sg(\T_i,\St_{i-1}))=\text{True}.$$
\end{definition}

If all state transitions are valid, then we say the ledger is a valid
ledger.

\begin{definition}
  $\Lg^{n} $ is a valid ledger iff
  $$\forall i\in[1,n],\ \Sv(\T_i, \St_{i-1}, \St_{i})=\text{True}$$
\end{definition}

\subsection{Modelling state sharing}\label{SEC:formal-state-sharing}

To formalize a framework for state sharing of permissioned ledgers, we
first distinguish between internal state and an external view of a
ledger $\Lg^{n}$. The internal state is shared by all internal parties
whereas the external view is the observation that external parties can
make about the ledger. The management committee is responsible for
maintaining both the internal state and the external view, and while
all internal parties see the same ledger, there can be multiple
different external views. This could be because the permissioned
network chooses to expose a partial view of its ledger to different
parties in the external world.

To prove the validity of a fact $F$ on a ledger $\Lg^{n}$, we
introduce the notion of an assertion. An assertion is a verifiable
proof against the internal state of the ledger that can be used to
validate $F$. To offer external parties a similarly verifiable proof
of facts against their view of the ledger, we need to adapt the
construct of an assertion to external views. We do this through the
notion of a wrapping function on an assertion. We use these three key
constructs of views, assertions and wrappings to formalize state
sharing in permissioned ledgers. Specifically, a proof of state system
for permissioned ledger $\Lg^{n}$ comprises three pairs of algorithms,
namely assertion generation and verification $(\Pg,\Pv)$, view
generation and verification $(\Vg,\Vv)$ and wrapping $(\WPg,\WPv)$. We
formally define and expound on each of these in this subsection.

For a given ledger, $\Lg^{n}$, we denote a digest $\D_i\in\D$ as a
unique representation of the state $\St_i$ and the set of facts in the
ledger as $\F_i\in\F$, where $\F_i=\T_i\cup \A_i\cup \M_i \cup
\Pl_i$. A verifiable assertion about any fact F is generated by the
management committee and is denoted as $\pi\in\Pi$. An assertion
offers a verifiable proof that a fact is included in a ledger
according to the policies of the management committee. For instance,
if the system states of $\Lg^{n}$ are validated by the management
committee by reaching an agreement via the
PBFT~\cite{castro1999practical} protocol, then the assertion $\pi$ can
be a proof of membership that an application state $A_{i,j}\in\A_i$ is
recorded on the ledger, signed by at least $2f+1$ management committee
members, where $f$ is the number of faulty nodes the system can
tolerate.

  \begin{definition}\label{def:pg-pv}
    Assertion generation and verification of ledger $\Lg^{n}$ on a
    fact is a pair of algorithms $(\Pg,\Pv)$ such that
    \begin{itemize}
    \item
      $\Pg: \F  \times \D \times \M \times \Pl \to \Pi$
    \item $\Pv: \F \times \Pi \times \D \to \text{True/False}$
    \end{itemize}
    and we have that
    $$\forall i\in[1,n], F\in\F, \Pv(F, \Pg(F,\D_i,\M_i, \Pl_i), \D_i)=\text{True}$$
  \end{definition}

  Since an assertion that is valid at state $S_i$ may not be valid at
  state $S_{i+1}$, internal clients should have the ability to use an
  assertion $\pi$ to not only verify that a fact is a part of the
  ledger, but also that it is correct at a given point in
  time. However, because clients do not have a direct view of states
  of a permissioned ledger, we need a mechanism to bridge the view
  between clients and the management committee. This is done through a
  deterministic function, $\tau(i)$ that maps a state $\St_i$ of the
  ledger $\Lg^{n}$ to a unique time point on a globally synchronized
  clock, where $i\in[1,n]$.

  \begin{definition}
    For a ledger $\Lg^{n}$, an assertion $\pi$ on a fact $F$ is
    \emph{valid} at state $\St_i$ iff
    $$\exists i\in[1,n], \Pv(F, \pi, \D_i)=\text{True}$$

    \noindent and we say $F$ is \emph{current} at time $t$ iff $F$ is valid at
    $\St_j$ such that
    $$\left\{\begin{array}{ll}{\tau(j)\leq t<\tau(j+1),} & {j\in[1,n-1]}, \\
              {\tau(j)\leq t} & {j=n.}\end{array}\right.$$
  \end{definition}

  To provide a verifiable proof to external clients, the management
  committee must release a commitment about the internal state of the
  ledger to the external world. This commitment, denoted as $V\in\V$,
  forms the external view and is defined through a pair of algorithms
  $(\Vg,\Vv)$.

\begin{definition}\label{def:vg-vv}
  View generation and verification of ledger $\Lg^{n}$ on a state is a
  pair of algorithms $(\Vg,\Vv)$ such that
    \begin{itemize}
    \item
      $\Vg: \D \times \M \times \Pl \to \V$
    \item $\Vv: \D \times \M \times \Pl \times \V \to \text{True/False}$
    \end{itemize}
    and we have that
    $$\forall i\in[1,n], V\in\V, \Vv(\D_i, \M_i, \Pl_i, \Vg(\D_i,\M_i, \Pl_i))$$
    $$=\text{True}.$$
\end{definition}

Since we can have multiple valid external views of a permissioned
ledger, we denote an external view $\V^k_i\in\V$ on $\St_i$ as the set
of $k$ possible individual views $V_i^1, \ldots, V_i^k$ on $\St_i$. In
practice, we expect that $\V_i^k$ is published by the management
committee. It is possible that $\Vg$ is not deterministic and there
exists multiple valid views on the same state. If all individual views
of a state are the same, i.e. $k=1$, then we say there exists a
consistent view of the ledger state.

\begin{definition}\label{def:one-view}
  The external view $\V^k_i$ on $\St_i$ of ledger
  $\Lg^n$ for some $i\in[1,n]$ is consistent if
  $$\forall x,y\in[1,k],\  V_i^x=V_i^y.$$

  \noindent If the external view is consistent for all $\St_i$, then we
  say the external view on $\Lg^n$ is consistent.
\end{definition}

Since a view may or may not be the digest of a state, the system should be
able to prove a fact against a released view. We define the proof of a
fact against a view as a wrapping of the proof of the fact
against the digest.

  \begin{definition}
    A wrapping for assertion generation $\Pg$ and verification $\Pv$ is
    a pair of algorithms $(\WPg,\WPv)$ such that
    \begin{itemize}
    \item $\WPg: \F \times \Pi \times \D \times \V \to \W$
    \item $\WPv: \F \times \W \times \V \to \text{True/False}$
    \end{itemize}
    and we have that $\forall i\in[1,n], F\in\F, \D_i\in\D, \M_i\in\M, \Pl_i\in\Pl$ and
    $$\Pg(F,\D_i,\M_i, \Pl_i)=\pi \land \Vg(\D_i,\M_i, \Pl_i)=V_i \Rightarrow$$
    $$\WPv(F, \WPg(F,\pi,\D_i, V_i), V_i)=\text{True}.$$
  \end{definition}

\subsection{Security properties}\label{SEC:state-sharing-security}

As addressed previously, the management committee may be \emph{trustworthy},
\emph{malicious but cautious}, or fully
\emph{malicious}. We do not consider the case of a trustworthy
management committee as the security guarantee of this case is fully
dependent on the consensus protocol of the permissioned ledger. We say
a system is \emph{Type-1} secure if it is secure against a malicious
attacker and \emph{Type-2} secure if it is secure against a malicious but cautious attacker.

\begin{definition}\label{def:type-1}
  A proof of state system for permissioned ledger $\Lg^n$ is Type-1
  secure iff $\forall$ $F\in\F, W\in\W, i\in[1,n], V_i\in\V$, we have
  that
  \begin{equation*}
  \begin{aligned}\WPv(F, W, V_i)=\text{True} &\Rightarrow&&\\
 \exists\D_i, \M_i, \Pl_i, \ \mbox{\it s.t.}~\ & \Pg(F, \D_i, \M_i, \Pl_i)=\pi \land &\\
 &\Vg(\D_i, \M_i, \Pl_i)=V_i \land &\\
 &\WPg(F, \pi, \D_i, V_i) = W.&
\end{aligned}
\end{equation*}

\end{definition}



A malicious but cautious committee will misbehave if its actions
cannot be detected. For a system to be resilient to such an adversary,
honest parties must have knowledge of the system that enables them to
detect and prove misbehavior. Different types of participants have
different sets of knowledge.  The internal client ($I$) has knowledge
of a set of facts $F$ and associated proofs $\pi$ through internal
inquiry, and may or may not have knowledge of the digest $\D_i$ of a
ledger state $\St_i$; the external client ($E$) has knowledge of a set
of $(F,W,V)$; and the members of the management committee ($M$) have
knowledge of $\Lg^n$. We use $K^X_A(d)$ to denote participants of type
$A$ having knowledge of some data $d$, such that the participants are
honest (if $X=H$) or rational (if $X=R$), where $A\in\{I,E,M\}$ and
$X\in\{H,R\}$. 

\begin{definition}\label{def:type-2}
  A proof of state system for permissioned ledger $\Lg^n$ is Type-2
  secure if and only if the following two conditions hold $\forall$
  $\F^*\subseteq\F, \W^*\subseteq\W, \V^*\subseteq\V$:
  \begin{itemize}
  \item $\forall F\in\F^*, W\in\W^*, V\in\V^*$, the system is Type-1 secure; and
  \item
    $\forall F\in \F\setminus\F^*, W\in\W\setminus\W^*, V_i\in\V\setminus\V^*$,
    $\WPv(F, W, V_i)=\mbox{\text{True}}$, we have that
    $\exists V'_i\in V^*, A\in\{I,E,M\}$ s.t.\ external view $V_i$ is not
    consistent with $V'_i$ and $K^H_A(V_i,V'_i)$.
\end{itemize}
\end{definition}

\section{Protocol}\label{SEC:protocol}

\begin{figure*}[!htb]
\includegraphics[width=0.95\textwidth]{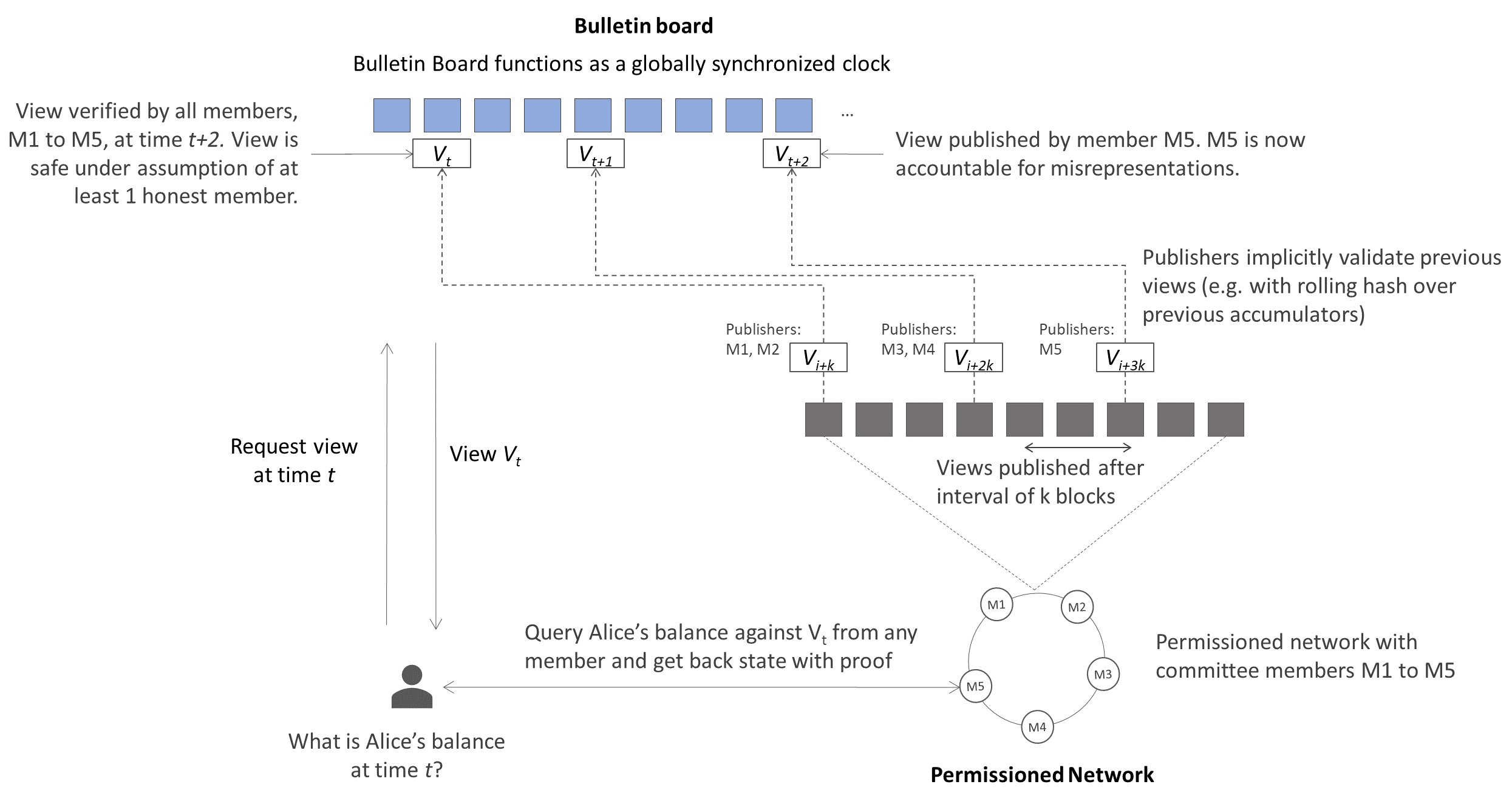}
\caption{Protocol overview: verifiable observations of state on
  permissioned ledgers. }
\label{fig:protocol-overview}
\end{figure*}

This section provides an overview of the proposed protocol and
parameters that influence its behavior. The protocol interfaces with
three entities: (i) the management committee of a permissioned network
(ii) an immutable public bulletin board, and (iii) external
clients. The protocol consists of four sub-components: (i) generation
and publication of views to the bulletin board, (ii) querying the
bulletin board for a view corresponding to time $t$, (iii) querying a
member of the management committee for a proof about facts in the
ledger against the view, and (iv) the verification of proofs by the
external client to establish the validity of facts. Two key parameters
that influence the behavior of this protocol are (i) the number of
members, $n$, of the management committee who choose to publish their
state digest at any give instance and (ii) the interval, $k$, at which
publications are made. These parameters represent trade-offs between
safety, liveness, and cost as discussed in Section
\ref{SEC:protocol_discussion}. Figure~\ref{fig:protocol-overview}
provides an overview of the protocol.

\subsection{Generation of State Digests}
\label{SEC:snapshot_generation} For every new block at height $i$,
each committee member computes a digest of the state $\St_i$ which
comprises the set of application states ($\A_i$), management committee
members ($\M_i$) and management policies ($\Pl_i$). The digest is
computed by updating an accumulator \cite{Benaloh1994}, a
cryptographic primitive that can represent a set of elements in the
form of a succinct digest and allows for proofs of membership and
non-membership without revealing other information about the
set. Various cryptographic accumulator schemes could be employed for
this purpose, ideally allowing for efficient dynamic updates.

Given that all nodes have full visibility of the state in the ledger,
and that the protocol offers deterministic finality
\cite{bano2019sok}, all members will compute the same digest for
$\St_i$. Nodes then create a view $\V_i$ using function $\Vg$ (as
defined in Def.~\ref{def:vg-vv}) that consists of the digest $\D_i$, a
signature over the digest and the corresponding block height $i$. A
view is therefore defined as $\V_i=(\D_i, i, \text{sig}_{\text{m}})$.

The protocol proceeds in rounds, wherein every $k$ blocks a new round
is initiated. In each round $r$, a subset of members from the
committee, $M^{r}_{i} \in \M_{i}$, can choose to publish their
accumulator at the current block height $i$, to the bulletin
board. The size of the subcommittee should at least be $f+1$ where $f$
is the failure threshold of the network, so as to ensure at least one
member publishes in each round. A network can choose different
strategies for publishing. Having all members publish in every round
can be expensive, but allows external clients to quickly determine if
a commitment has been seen by all members and is therefore safe. A
deterministic selection of unique subsets of members on every round,
such as a simple round robin assignment, trade-offs cost for delayed
safety. See Section~\ref{SEC:protocol_discussion} for further
discussion on the implications of selection strategies in relation to
safety and liveness.


\subsection{Publishing Commitments}\label{sec:publish-commitments}

Publishing views to an immutable bulletin board provides visibility to
external clients and mitigates the potential for a network to
equivocate on facts about the state of the ledger with external
clients. The bulletin board uses a smart contract published by the
permissioned network that is known to external clients \textit{a
  priori}. The smart contract is bootstrapped with a policy specifying
a list of public keys of the management committee. We assume
establishing the identity and validity of public keys occurs
off-chain, coordinated between external clients and management
committees. The bulletin board performs functions that include
tracking published views for accountability, ensuring that views do
not conflict, and offering a mechanism by which parties can report
conflicts.

Appendix \ref{appendix:bulletin-board-algorithm} presents an algorithm
describing how the bulletin board works. A publication to the bulletin
board is a tuple $(V^n_i, H^n_i)$ where $V^n_i$ is the view at $L_i$
computed by subcommittee member $n$, and $H^n_i$ a rolling hash of all
past views observed by $n$, up to $\V_i$. The rolling hash of views is
trivially computed by each node on every new view generation by
concatenating the hash of the previous accumulator with the previous
rolling hash, and hashing the result. The rolling hash offers two
benefits: (i) it serves as a mechanism by which parties implicitly
validate views from previous rounds in which they did not participate
(ii) it adds further accountability for the publishing party. If in a
previous round another member published a false accumulator, and
another member in a subsequent round provided a rolling hash which
included that false accumulator, it is evident that the parties are
colluding and can be held accountable.

\subsection{Detection and Reporting}


There are two scenarios in which an invalid commitment could
be detected by a single honest management committee member $m_{h}$:
\begin{enumerate}[i]
\item $m_{h} \in M^{r}_{i}$: The honest member is part of the
  subcommittee publishing in the round. In this case, the member's
  accumulator will conflict with accumulator values submitted by
  malicious committee members or in the rolling hash of past
  accumulators. This inconsistency is detected by the bulletin board.
\item $m_{h} \notin M^{r}_{i}$: The honest member is not publishing in
  the current round but observes events of publications by other
  members. If the member detects that a posted view does not match its
  accumulator for the ledger height, it reports the conflict to the
  bulletin board.
\end{enumerate}

The delay in detecting and reporting conflicting commitments posted by
a malicious member 
is the time before an honest member can detect and report view
discrepancies. In the second model where members observe events from
the bulletin board, this time is only influenced by network delays and
crash failures.

The detection of a conflict and its reporting on the bulletin board
enables the rest of the management committee and external clients to
become aware of the presence of malicious behavior. However, such
conflicts don't provide any information about the validity of a view,
and by extension, which parties were malicious and which ones were
honest. The process of resolving this dispute is to be handled
out-of-band by the respective committee members. In practice, this
could involve an external arbitrator.

\subsection{State Query and Verification}
External clients use the bulletin board as a source of views of the
permissioned ledger. An external client will query the smart contract
for views either seen by all members or views that were recently
published. The bulletin board's block height allows the client to
reason about the time at which publications were made. The client will
use the view from the bulletin board to query a committee member for a
given state (e.g. an application state $A_{i,j} \in \St_i$). The node
will use the $\Pg$ algorithm to create an assertion $\pi$ that the
application state $A_{i,j}$ is in the state set $\St_i$. $\pi$
includes the membership proof of the state in the accumulator.

\subsection{Discussion}\label{SEC:protocol_discussion}
The proposed protocol offers guarantees of detectability of malicious behavior
associated with state sharing and accountability of members
responsible for the misbehavior, assuming at least one honest member
in the network. In a permissioned context, this event would have legal
and reputational ramifications that would serve as a deterrent for
misbehavior by a malicious but cautious adversary.

However, a safety guarantee for the external client on a state
commitment is only achieved when all members of the committee have
implicitly or explicitly published the same accumulator for a given
block height (see~\ref{sec:publish-commitments}). That is, it is
explicit when it is a member of the subcommittee publishing in a given
round, or implicit when, in a future round, it endorses a history of
past accumulators of which that state commitment is a part. This
however, affects liveness if the honest member is unavailable. More
practically, an external agent can utilize a commitment without
requiring a guarantee that all members have agreed on the same
accumulator value, by relying on the fact that parties will be held
accountable for misrepresenting state. For instance, an external agent
that does not expect any collusion between parties can utilise a
commitment as soon as it is published. Similarly, an agent that does
not expect more than one third of nodes to collude can wait until more
than a third of the members have published the same accumulator for a
given block height. More generally, an external agent that assumes a
colluding set of $c$ members in the network would need to wait for
$c+1$ votes on a commitment.

The guarantee on currency of state is influenced by the intervals at
which state commitments are published, expressed as the number of
blocks $k$. Reducing $k$ increases the frequency of commitments and
thus offers external clients better state coherence with the
permissioned ledger. While this is desirable, it increases the
overhead and cost of publishing commitments. In addition, this could
also be constrained by the throughput of the bulletin board.

\section{Proof of Concept with Hyperledger Fabric}\label{SEC:proof-of-concept}

Hyperledger Fabric is an open-source permissioned blockchain designed
for enterprises \cite{androulaki2018hyperledger}. Fabric's design uses
an execute-order-validate model for appending transactions to the
blockchain. A Fabric network consists of an ordering service and one
or more channels. Each channel consists of a set of peers, managed by
independent organizations, and represents a unique ledger with its own
genesis block. The ledger is based on the account model. A subset of
peers run smart contracts and endorse transaction proposals submitted
by clients. All peers validate blocks from the ordering service. We
refer the reader to Fabric's protocol documentation~\cite{Fabric} for
details.

\subsection{Enabling state sharing in Hyperledger Fabric.}
We describe the implementation of the protocol presented in
Section~\ref{SEC:protocol} for sharing state from a Fabric network
to an external client. Fabric agents listen to block
events~\cite{fabricevents} from peers and maintain a digest of the
entire ledger state in the form of an RSA accumulator \cite{Benaloh1994}. They publish
the digest as a view to a contract on the Ethereum
blockchain. External clients fetch views from the Ethereum contract
and make a request for a state corresponding to the view from the
Fabric agent. The agent retrieves the state from the Fabric ledger and
generates a proof of membership of that state in the accumulator. The
external client verifies the state using this proof.

\myparagraph{Cryptographic accumulators for view generation.}  Our
design uses a dynamic RSA accumulator \cite{Benaloh1994,Camenisch2002}
to represent a succinct digest of the Fabric ledger. We make use of an
open-source library from Westar Labs~\cite{westarlabs2019}. The RSA
accumulator is set up with an RSA modulus $N$ that is the product of
two safe primes $p$ and $q$. A base generator $g$ is selected from the
set of quadratic residues modulo $N$. In order for each Fabric agent
to create the same RSA accumulator, they use identical seeds to
deterministically generate $p$, $q$ and $g$. For details on the
construction of the accumulator and membership proofs see Appendix
\ref{appendix-accumulator}.

\myparagraph{Peer agents as generators of views.} Each Fabric agent
maintains an RSA accumulator as a digest of the complete state of the
ledger using the $DGen$ function described in Algorithm~\ref{DGen} in
Appendix~\ref{appendix-accumulator}. To do this, the agent subscribes
to all block events from the peer, iterating through valid
transactions and creating a list of the key/value pairs that are
updated in the ledger, denoted the $KV$ write set
\cite{FabricKVProto}. Each $KV$ write is converted to a prime through
the PrimeGen function as described in Appendix
\ref{appendix-accumulator}. If the $KV$ write is marked with an
``isDelete" flag, the element is deleted from the accumulator,
otherwise it is added.

For each block height $i$, the Fabric peer agent also maintains a rolling hash, $H_i$, of
the history of accumulators. The rolling hash is computed by hashing
the concatenation of the previous rolling hash with the previous
accumulator. The rolling hash enables the efficient verification of
consistency of the accumulator between peer agents. For all peer
agents to have calculated the same rolling hash for a block height
they must have the same accumulator value for every previous
block. Both the accumulator and the rolling hash are stored in a local
persistent store of the Fabric agent. The agent storing prior accumulators for
every block height allows proofs to be generated based on historical
state.

\myparagraph{Bulletin board.} For a block height $i$, the agent
publishes the accumulator, the rolling hash, and its signature to the
Ethereum contract. We use Ethereum as the bulletin board for its
desirable properties, which include a tamper-resistant log for
non-repudiability, public availability, and support for smart
contracts. A $ManagementCommittee$ contract maintains a list of the
public keys of agents of Fabric peers that make up the management
committee. These public keys are recorded against the corresponding
Ethereum address of the agent. The $LedgerState$ contract accepts
views submitted by Fabric agents. The contract only accepts views if
the transaction signature corresponds to a valid Ethereum account
recorded in the $ManagementCommittee$ contract.

\myparagraph{Proof generation and verification of facts.} An external
client obtains a view from the LedgerState contract by specifying a
block number of the Ethereum chain. It uses the view to request a
Fabric agent for a fact about the state of the ledger corresponding to
that view, e.g. a key/value pair of some application state
$A_{i,j}$. The Fabric agent queries the peer for all historical states
of the application and finds the latest version that is present in the
accumulator sent by the external client. The agent then generates a
membership proof of the prime representation of that state in the
accumulator. The external client verifies this membership proof in
constant time.

\subsection{Security analysis.} Loosely speaking, assuming the
bulletin board guarantees consensus safety, then all honest
participants, including members of the management committee and
external clients, have a consistent view (as in
Def.~\ref{def:one-view}) of the same internal state. 
A misbehavior can lead to conflicting valid view of the same block height.
In addition, the correctness and
soundness of the proof generation and verification is guaranteed by
the underlying accumulator. When the management committee tries to add
a false commitment to the bulletin board, such behavior will be
detected by any honest party with the knowledge of the correct
commitment at the same block height. So, any honest party with the
knowledge of a conflict can raise an alert on the misbehavior with
evidence. This prevents attacks from a malicious but cautious
committee.

\begin{theorem}\label{theorem:1}
  Under the strong RSA assumption, with a secure signature scheme, and
  the security of the bulletin board, the permissioned ledger $\Lg^n$
  of the proposed protocol is Type-2 secure.
\end{theorem}

Proof of Theorem~\ref{theorem:1} can be found in Appendix~\ref{appendix:proof}.

\section{Performance Evaluation}\label{SEC:perf-eval}

We compared the performance of our protocol with a Fabric network to
show whether the Fabric commitment agent can keep up with the peer
network and publish \textit{current} state to the bulletin board as
the ledger grows.  We used maximum throughput (transactions per
second, or TPS) and latency as evaluation metrics.  Studies have
revealed how Fabric performs in different
conditions~\cite{thakkar2018performance} and provided guidelines to
engineer a Fabric network for high performance~\cite{Ferris2019perf}.

\subsection{Experimental Setup}\label{SEC:expt-setup} 
Since we wish to compare the processing ability of our protocol
relative to that of Fabric, any Fabric setup will suffice to produce
an estimate as long as both the base network and our protocol are
measured in identical environments.  We conducted tests on a single VM
running a Fabric 2 network in Docker containers, using Hyperledger
Caliper~\cite{Caliper} for workload generation and throughput
measurement.  The VM was allocated 32 vCPUs (Intel(R) Xeon(R) Gold
6140 CPU @ 2.30GHz) and 64GB of memory.  The Caliper installation was
bound to Fabric version 2.1.0, running on the same VM as the Fabric
network.  We used two different networks for our experiments: a
single-peer network running the Smallbank benchmark~\cite{Alomari2008}
and a more production-like three-organization network running a trade
contract~\cite{Trade-Network}. The source code is available
online\footnote{https://github.com/dlt-interoperability}.

Details of the two sets of networks and benchmark applications are as
follow:
\begin{enumerate}
    \item \textbf{Basic Smallbank}: This is the simplest possible Fabric
    network, with a single peer (using Go LevelDB) and a certificate authority
    (CA) in a single organization, and a single ordering node running in
    \textit{solo} mode. We used the \textit{smallbank} benchmark application
    \cite{Alomari2008} offered by Caliper, which simulates common bank
    transactions. The \textit{create\_account} function was selected to drive
    transaction load and the \textit{query} (account balance) function for
    query load.
    \item \textbf{Trade}: This is a more realistic network, with three peers
    (using CouchDB) and CAs across three organizations (representing exporter,
    importer, and regulator) and a five-node Raft ordering service
    \cite{Trade-Network}. This network runs a \textit{trade} contract that
    supports trade requests, acceptances, and views \cite{Trade-Contracts}. The
    \textit{requestTrade} function was selected to drive transaction load and
    the \textit{GetTradeStatus} function for query load.
\end{enumerate}

For each network and benchmark set, we first determined the maximum throughput
achievable on our test machine without the state commitment protocol components,
by:
\begin{itemize}
    \item Launching the network with Docker Compose and installing the benchmark
    contract, and
    \item Increasing the transaction submission rate in the Caliper workload
    configuration until the TPS fell instead of increasing (indicating
    saturation).
\end{itemize}

For each network configuration, transaction arguments were chosen to
guarantee that there were no conflicting transactions within a
block. Conflicts can occur when a transaction records it has read
state from the ledger and a prior transaction in the block has altered
that state. To avoid conflicts, our configuration specified that keys
were only created and never updated. 

\subsection{State Commitment Capacity}\label{SEC:throughput}
First, each Fabric network was launched without our protocol and its
baseline performance (maximum TPS) measured. Each was then restarted
with protocol components, namely one Fabric agent (instrumented to
record TPS and latencies) for each peer and an Ethereum test network
as a bulletin board. We ran tests using the maximum throughput
workload configurations. Table~\ref{tab:tps} shows the baseline
platform and protocol throughputs, which are measures of their
respective processing capacities. (For Trade, the BPS and TPS are the
minimum of the three agents' reported values.)

\begin{table}[htb]
    \caption{Transaction Processing Capacity}
    \label{tab:tps}
    \centering
    \begin{tabular}{|c|c|c|c|c|}
    \hline
    \textbf{Scenario} & \textbf{Basic Smallbank} & \textbf{Trade} \\ \hline
    \textbf{Num Tx} & 40,840 & 13,240 \\ \hline
    \textbf{Max Network TPS} & 370 & 138 \\ \hline
    \textbf{Protocol BPS} & 27.8 & 12.1 \\ \hline
    \textbf{Protocol TPS} & 278.4 & 121.5 \\ \hline
    \end{tabular}
\end{table}

The maximum throughput achievable on our test machine using the basic
network configuration and benchmark application was 370 TPS. The
commitment agent was able to achieve a maximum throughput of 278 TPS
in comparison. (\textit{Note}: each block contains the maximum of 10
transactions as configured; hence the TPS is 10 times the blocks per
second, or BPS). This includes the overhead of publishing to the
bulletin board every 120 blocks for the Smallbank application, which
corresponds to roughly every 4.3 seconds. The gas cost of publishing
each commitment to the Ethereum bulletin board was 203,514.

\subsection{Query Capacity}\label{SEC:query-perf}
To understand how responsive an agent is to queries from external
clients, we performed multiple state and proof queries after every
batch of 120 blocks. The average query response time observed by an
external client was around 231ms. This included the overall process of
first querying the Ethereum bulletin board for a commitment, then
querying the Fabric agent for a state that corresponds to the
commitment and finally verifying the returned proof of membership.
This time was independent of the number of states that had been
accumulated.

\subsection{Analysis and Improvements}\label{SEC:improvements}
Our measurements revealed that the agent was roughly 24.8\% slower
than the peak TPS for the simplest benchmark application, Smallbank,
and 12.3\% slower than the maximum TPS for the Trade
application. While these results suggest that under a high transaction
volume the currency of published data cannot be guaranteed, in
real-world enterprise networks we are likely to encounter networks and
contracts that are much more complex than these benchmarks. In such
scenarios, transactions of interest would occur at lower frequencies
and hence it is likely that our agents will deliver adequate
performance in practice. In addition, we found that different agents
deployed for each organization in the Trade network delivered similar
TPS, confirming that size of the management committee and the network
have no impact on agent throughput.

Several improvements can be made to the implementation and
configuration of the protocol to significantly enhance its
performance. From an implementation perspective, basic optimisations
can be made to the agent to reduce I/O and compute overheads. In
addition, the configuration of the protocol can be relaxed depending
on real world requirements. For instance, some use cases may require
less frequent publication of accumulators. Similarly, networks could
also choose to only compute and publish digests for a subset of keys
that are shared with external clients, as opposed to the entire
ledger. Such variations can reduce the cost and the TPS requirements
on the agent. Further, to optimise query performance, frequently
queried keys and their associated proofs can be cached.

The commitment scheme employed in the protocol could also be
improved. For instance, by consolidating the update of multiple keys
into a single operation \cite{Boneh2019}, the cost of updating
accumulators can be significantly reduced. In addition, other
commitment schemes \cite{agrawal2020kvac, zhangtick} could also be
investigated in future work.

\section{Conclusion}\label{SEC:conclusion}

This paper addressed an important problem challenging external
consumers of state of permissioned ledgers: verifying the validity of
state and reasoning about its currency, assuming at least one honest
member in the network. This work lays the foundation for enabling
cross-chain communication across permissioned networks, as well as
integrating permissioned networks with legacy enterprise
applications. We provided a formalization of the problem of state
sharing under different adversarial conditions and proposed the design
of a protocol for state sharing with safety and liveness
trade-offs. We also presented a security analysis supporting our
protocol design. Our experiments and analysis with Hyperledger Fabric
showed the viability of the protocol with a number of avenues for
improving performance.

\section*{Acknowledgement}
This work was partially supported by the Australian Research Council
(ARC) under project DE210100019. The authors would like to thank Yacov
Manevich for his review and feedback.

\bibliographystyle{splncs04}

\bibliography{ref}

\appendices

\section{Related Work}\label{sec:rw}

\myparagraph{Light Clients.} Verifying blockchain states without
having full access to the ledger was envisioned in the original
Bitcoin paper ~\cite{nakamoto2019bitcoin}. For instance, resource
constrained light clients may need to confirm the inclusion of their
transactions in the blockchain, but lack the resources to store and
validate the entire ledger. Bitcoin leverages the Simplified Payment
Verification (SPV) technique~\cite{nakamoto2019bitcoin} to allow light
clients to maintain only block headers and receive Merkle proofs from
full nodes about transactions related to them.  However, there are a
number of limitations to this protocol. Light clients have to validate
and store every block header, trust full nodes for transaction
validation, and divulge significant information and cede privacy.
There have been numerous efforts to address these
limitations~\cite{PoPoW,kiayias2017non,FlyClient,PoNW,zhangtick,Hearn2012,Gervais,BITE}.
Nonetheless, SPV-based schemes are not applicable to permissioned
ledgers~\cite{Fabric,Corda,Besu,Quorum,MultiChain} as external clients
do not have visibility on the ledger's history, including block
headers and management policies.

\myparagraph{Layer-2 Protocols.} Layer-2 protocols are approaches to
scaling that allow users to transact off-chain, enabling lower
latencies, little or no fees and private transactions. Layer-2
protocols, sometimes referred to as
commit-chains~\cite{khalil2018commit}, are non-custodial and rely on
the parent chain for security. Examples of these systems include
Plasma~\cite{ethhubplasma}, NOCUST~\cite{khalil2018nocust}, Optimistic
Rollups schemes such Arbitrum~\cite{kalodner2018arbitrum} and
ZK-Rollups schemes such as zkSync~\cite{matterLabs,zksync}. These
protocols checkpoint state at periodic intervals to Ethereum and rely
on a challenge-response game for integrity and/or data
availability. In the case of Plasma or Optimistic Rollups users submit
fraud proofs to a designated contract when fraud is detected. In the
case of ZK-Rollups, consistency is inherent in the zero-knowledge
proof construction, but operators are still challenged on data
availability. Third-party watchers such as PISA~\cite{mccorry2019pisa}
can augment Layer-2 protocols by being incentivized to monitor the
actions of operators. In contrast to the model in which Layer-2
protocols operate, which includes non-custodial services with safe
exits to the parent chain, permissioned networks are closed systems
which create and manage their own state. External clients, including
third-party watch services, are unable to monitor the actions of the
network since no data is revealed for privacy and confidentiality
reasons. Furthermore, the general purpose nature of these ledger
technologies and their broad applications to real world problems makes
using zero-knowledge schemes such as ZK-SNARKS~\cite{ben2013snarks},
ZK-STARKS~\cite{ben2018scalable} and
Bullet-Proofs~\cite{bunz2018bulletproofs} difficult and expensive.

\myparagraph{Cross-Chain Communication Protocols.} Protocols such as
Cosmos~\cite{Cosmos}, Polkadot~\cite{Polkadot},
Cardano~\cite{cardano}, Ren~\cite{ren} and BTC Relay~\cite{btcrelay}
enable communication of state, representing either data or assets,
across different chains. These systems rely on an intermediate network
of nodes in order to relay messages. In order to ensure that the state
communicated is valid, these intermediaries require access to the
internal state of the source chains (even if only block
headers). Systems such as Cosmos and Polkadot also require chains to
follow prescribed protocol specifications (either directly or through
bridges) in order to participate in the ecosystem. However, the number
of real-world deployments of permissioned ledgers that already exist
make it impractical to migrate them into ecosystems like Cosmos and
Polkadot. Furthermore, permissioned ledgers are closed sovereign
networks that don't reveal any chain data externally, making them
difficult to integrate.

Recently, efforts have emerged around relaying states across
permissioned ledgers. Abebe et. al~\cite{Abebe2019} and Hyperledger
Cactus~\cite{Cactus2020} describe systems for state sharing between
independent permissioned networks by relaying attestations. While
these systems have practical value, they rely on the external client
placing significant trust on the committee members signing the state
for its validity and currency. Our work, however, assumes a worst case
scenario that requires only one honest member in the committee for
providing guarantees on the validity of state while allowing clients to
reason about its currency.

\section{Bulletin Board}\label{appendix:bulletin-board-algorithm}
Pseudocode of the bulletin board is presented in algorithm
\ref{alg:BB}.  The bulletin board records commitments from any
management committee member, $m$, about the state of a permissioned
ledger at height $i$ (in the form of a view $\V^m_i$), and its lineage
thus far (in the form of a rolling hash $H^m_i$). Each time a
commitment is submitted, the bulletin board checks that the value of
the submitted commitment does not conflict with what has already been
submitted by other members, and that the member's history of past
accumulators matches the bulletin board's own up to that point. If
either of these conditions fail, the bulletin board raises an event to
notify all members of this inconsistency, as it indicates malicious
behavior. The list of all members that have explicitly published a
commitment for ledger height $i$ is stored, and their rolling hash is
observed as an endorsement of past commitments. Using these, the
bulletin board knows which past views have been explicitly and
implicitly endorsed by all members of the committee. The bulletin
board also provides a mechanism for committee members to report
conflicts that they observe at any point. The bulletin board notifies
all management committee members and subscribed external clients of
events such as conflicts being detected and new commitments being
published so as to ensure detectability and accountability of actions.

\begin{algorithm}
    \caption{Bulletin Board Publishing and Conflict Reporting}
    \label{alg:BB}\small
\begin{algorithmic}
 \Require $\M_i, \Pl_i$
 \Function{publishView}{$i$, $m$, $\V^m_i$, $H^m_i$, $sig_m$}
 	\State \Verify{$m$ is a committee member and signature is valid}
 	\State $V_i, H_i \leftarrow$ \textproc{getCommitmentFor($i$)} \Comment{view already published for $i$ if any}
 	\If{$\V_i$ \textbf{and} ($\V_{i}$ != $\V^m_i$ \textbf{or}  $H_{i}$ != $H^m_i$)} 
		\State \textproc{reportViewConflict($i$, $m$, $\V^m_i$, $H^m_i$, $sig_m$)}
	\ElsIf{$!\V_i$ \textbf{and} $H^m_i$ != $computeRollingHash(i, V^m_i)$ }
		\State \textproc{reportViewConflict($i$, $m$, $\V^m_i$, $H^m_i$, $sig_m$)}
	\Else
		\State \textproc{recordView($i$, $m$, $\V^m_i$, $H^m_i$, $sig_m$)}
		\State emit event ViewPublished($i$, $m$, $\V^m_i$)
  \EndIf
\EndFunction

\Function{reportViewConflict}{$i$, $m$, $\V^m_i$, $H^m_i$, $sig_m$}
 	\State \Verify{$m$ is a committee member and signature is valid}
 	\State $V_i, H_i \leftarrow$ \textproc{getCommitmentFor($i$)} \Comment{view already published for $i$ if any}
 	\If{$\V_i$ \textbf{and} ($\V_{i}$ != $\V^m_i$ \textbf{or}  $H_{i}$ != $H^m_i$)} 
		\State $V_i \rightarrow State \leftarrow "DISPUTED"$
		\State emit event ViewCommitmentConflict($i$, $\V_i$, $H_i$, $\V^m_i$, $H^m_i$)
	\EndIf
\EndFunction
\end{algorithmic}
\end{algorithm}

\section{RSA Accumulator} \label{appendix-accumulator}

 \begin{algorithm}
    \caption{Digest generation function $DGen$}\small
    \label{DGen}
    \begin{algorithmic}
      \Input{$\T_i$, $\St_{i-1}$, $z_{i-1}, g$}
      \Output{$z_i$}
    \end{algorithmic}
    \begin{algorithmic}
      \For {all $T_{i,j} $ in $ \T_i$} {
        \If {$T_{i,j}$ is valid} {
          \For {all $KV_k^i$ in $T_{i,j}$} {
            \If {$KV_k^i$ contains isDelete flag}
            \State $z_i=g^{\St_{i-1}-KV_k^i}\text{ mod }N$
            \Else
            \State $z_i=z_{i-1}^{PrimeGen(KV^i_k)}$ mod $N$
            \EndIf
          }
          \EndFor
        }
        \EndIf
      }
      \EndFor
      \State $z_i$
    \end{algorithmic}
\end{algorithm}

An RSA accumulator is a cryptographic primitive first described by Benaloh and
de Mar \cite{Benaloh1994}. It generates a succinct digest of a set of elements
and allows for proof of membership of an element in the set without revealing
any other information about the set. Variations of this accumulator to provide
dynamic updates and both membership and non-membership proofs were later
proposed \cite{Camenisch2002,Li2007,Baldimtsi2017}. The accumulator is set up
with an RSA modulus $N$ that is the product of two safe primes $p$ and $q$. A
base generator $g$ is selected from the set of quadratic residues modulo $N$.
The accumulated value for a set $X = \{x_1, ..., x_n\}$ is defined as $z =
g^{x_1...x_n} \text{ mod } N$. Importantly, this set must consist of primes and
therefore any non-primes must be converted to primes through a PrimeGen
conversion function. In our implementation, the PrimeGen function hashes the
element to be accumulated with an incremental nonce and checks the result for
primality until a suitable nonce is found. However, there are other methods for
prime conversion, for example in \cite{Baric1997}.

Adding a new element, $x'$, to the accumulator is done by $z'=z^{x'} \text{ mod} 
N$, where $z'$ is the new accumulated value equivalent to $g^{X\cup\{x'\}}$.
Deleting element $x'$ from the accumulator can be done efficiently by using
auxiliary information about the accumulator, namely $\Phi(N)=(p-1)(q-1)$.
Removal of $x_j$ from the accumulator using this method is done by
$z'=z^{x_j^{-1} \text{ mod }\Phi(N)}$ mod $N$, where $z'$ is the accumulated
value equivalent to $g^{X\backslash\{x_j\}}$.

A membership proof for an element $x_j$ in $z$ can similarly be generated using
the accumulator's auxiliary information to calculate $z^{x_j^{-1} \text{ mod
}\Phi(N)}$. The consumer of the proof can verify it in constant time by checking
that $w^{x_j} \text{ mod } N=z$. A description of the generation of a
non-membership proof that can be generated efficiently and verified in constant
time is presented in \cite{Li2007}.

The dynamic accumulator scheme also allows update of membership proofs in
constant time. After adding a new element, $x'$, update of the witness $w$ of
element $x$ in $z$ to a witness that proves $x$ is a member of $z'$ is done by
$w'=w^{x'}$. After deletion of an element $x'$, the witness $w$ of $x$ in $z$
(where $x\neq x'$) can be updated to prove $x$ in $z'$ by $w'=g^bz'^a$, where
$a$ and $b$ are integers computed through the extended Euclid algorithm such
that $ax+bx'=1$. The efficient update of a non-membership witness is described
in \cite{Li2007}.

As the elements included in the accumulator must be primes, every key must find
a suitable nonce such that the resulting hash of the key and its nonce is prime.
Finding this nonce can be done in a deterministic fashion. However, finding this
nonce takes some computational effort and therefore the RSA accumulator in our
implementation stores a map of all the accumulated keys and their nonces. This
enables fast lookup of whether an element is in the accumulated set and the
ability to create the accumulator after deletion of a key from the ledger,
without having to retrieve all state from the peer's ledger.

\section{Proof of Theorem~\ref{theorem:1}}\label{appendix:proof}
\setcounter{theorem}{0}

\begin{theorem}
  Assuming the strong RSA assumption, a secure signature scheme, and
  the security of the bulletin board, the permissioned ledger $\Lg^n$
  of the proposed protocol is Type-2 secure.
\end{theorem}

\begin{IEEEproof}
  We prove this by contradiction. Assume that the system is not Type 2
  secure. Then, according to Definition~\ref{def:type-2}, either of
  the following two conditions hold: (1) there exists
  $F\in\F^*, W\in\W^*, V\in\V^*$, such that the system is not Type-1
  secure; or (2) there exists
  $F\in \F\setminus\F^*, W\in\W\setminus\W^*, V_i\in\V\setminus\V^*$,
  such that $\WPv(F, W, V_i)=\mbox{\text{True}}$, but there does not exist
  $V'_i\in V^*$ or $A\in\{I,E,M\}$, s.t.\ external view $V_i$ is not
  consistent with $V'_i$ or $K^H_A(V_i,V'_i)$ does not hold.

  We first prove condition (2) by contradiction.  For all
  $F\in \F\setminus\F^*, W\in\W\setminus\W^*, V_i\in\V\setminus\V^*$, if
  $\WPv(F, W, V_i)=\mbox{\text{True}}$ and the system is not Type-1 secure,
  then according to Definition~\ref{def:type-1}, either of the
  following three subconditions does not hold:
  \begin{enumerate}
  \item $\Pg(F, \D_i, \M_i, \Pl_i)=\pi$
  \item $ \Vg(\D_i, \M_i, \Pl_i)=V_i$
  \item $ \WPg(F, \pi, \D_i, V_i) = W$
  \end{enumerate}

  If the subcondition 1, i.e., $\Pg(F, \D_i, \M_i, \Pl_i)=\pi$, does
  not hold, then there exists a proof $\pi'$ such that
  $\Pv(F, \pi', \D_i)=\text{True}$ and there does not exist
  $\Pg(F, \D_i, \M_i, \Pl_i)=\pi'$ for some $F$, $\D_i$, and
  $\M_i$. Assuming the strong RSA assumption, this contradicts the
  soundness of the RSA accumulator, due to Theorem 2 of
  \cite{Camenisch2002}.

  If subcondition 2, i.e., $ \Vg(\D_i, \M_i, \Pl_i)=V_i$, does not
  hold, then the view $V$ is a fake view for an external party. In
  other words, $(\D_i, \M_i, \Pl_i)$ does not represent the system
  state $\St_i$. One condition for the validity of the view is the
  wrapped view must be present in the bulletin board. As we will
  analyse the wrapping process in subcondition 3, let's assume that
  the wrapping process is secure. In addition, as we assume the
  security of the underlying bulletin board, any party has access to
  the immutable record of it. According to our security model, there
  exists at least one honest party which has access to the bulletin
  board. Thus, this party will have the knowledge of this fake view
  represented by $(\D_i, \M_i, \Pl_i)$ such that
  $\Vv(\D_i, \M_i, \Pl_i, V_i)=\text{False}$, and the corresponding correct view
  $V'_i$. So, $K^H_A(V_i,V'_i)$ would hold such that external view
  $V_i$ is not consistent with $V'_i$. This is evidence of
  misbehavior.

  If subcondition 3, i.e., $ \WPg(F, \pi, \D_i, V_i) = W$ does not
  hold, then either the proof $\pi$ is forged, or the view $V_i$ is
  forged, or the verification $\WPv(F, W, V_i)$ is not sound. The
  first two subconditions are already proved. For the last one, as
  the wrapping generation and verification in our proposed scheme is
  done by using a secure signature scheme. If $\WPv(F, W, V_i)=\text{True}$
  and $ \WPg(F, \pi, \D_i, V_i) = W$ is not true, then this contracts
  to the unforgeability of the signature scheme. Thus, this
  contradicts the assumption of a secure signature scheme.

  With the above proofs w.r.t. the three subconditions, condition (2)
  does not hold by contraction. Condition (1) has similar
  subconditions that are required for proving Condition (2), and the
  proof is similar to the verification of condition (2).
\end{IEEEproof}

\end{document}